\documentclass{pasj00}

\begin{document}
\SetRunningHead{H. Uchiyama et al.}{CTI correction for the Spaced-row Charge Injection of the XIS}
\Received{2000/12/31}
\Accepted{2001/01/01}
\title{New CTI Correction Method for the Spaced-Row Charge Injection 
of the Suzaku X-Ray Imaging Spectrometer}
\author{
Hideki \textsc{Uchiyama}\altaffilmark{1}, 
Midori \textsc{Ozawa}\altaffilmark{1}, 
Hironori \textsc{Matsumoto}\altaffilmark{1},\\ 
Takeshi Go \textsc{Tsuru}\altaffilmark{1}, 
Katsuji \textsc{Koyama}\altaffilmark{1}, 
Masashi \textsc{Kimura}\altaffilmark{2}, \\
Hiroyuki \textsc{Uchida}\altaffilmark{2}, 
Hiroshi \textsc{Nakajima}\altaffilmark{2}, 
Kiyoshi \textsc{Hayashida}\altaffilmark{2},\\ 
Hiroshi \textsc{Tsunemi}\altaffilmark{2}, 
Hideyuki \textsc{Mori}\altaffilmark{3}, 
Aya \textsc{Bamba}\altaffilmark{3}, \\
Masanobu \textsc{Ozaki}\altaffilmark{3}
Tadayasu \textsc{Dotani}\altaffilmark{3}, 
Dai \textsc{Takei}\altaffilmark{4},\\
Hiroshi \textsc{Murakami}\altaffilmark{4},
Koji \textsc{Mori}\altaffilmark{5},
Yoshitaka \textsc{Ishisaki}\altaffilmark{6},\\
Takayoshi \textsc{Kohmura}\altaffilmark{7},  
Gregory \textsc{Prigozhin}\altaffilmark{8}, 
Steve \textsc{Kissel}\altaffilmark{8}, \\
Eric \textsc{Miller}\altaffilmark{8}, 
Beverly \textsc{LaMarr}\altaffilmark{8}, 
and Marshall \textsc{Bautz}\altaffilmark{8} }

\altaffiltext{1}{Department of Physics, Graduate School of Science,
Kyoto University,\\ Sakyo-ku, Kyoto 606-8502}
 \email{E-mail (HU) uchiyama@cr.scphys.kyoto-u.ac.jp}
 \altaffiltext{2}{
Department of Earth and Space Science, Osaka University,\\ 
Machikane-yama, Toyonaka, Osaka 560-0043}
 \altaffiltext{3}{Institute of Space and Astronautical Science,
Japan Aerospace Exploration Agency,\\ 
3-1-1 Yoshinodai, Sagamihara, Kanagawa 229-8510}
\altaffiltext{4}{Department of Physics, Rikkyo University,\\ 
3-34-1 Nishi-Ikebukuro, Toshima-ku, Tokyo 171-8501}
\altaffiltext{5}{Department of Applied Physics, University of Miyazaki,\\ 
1-1 Gakuen Kibana-dai Nishi, Miyazaki 889-2192}
\altaffiltext{6}{Department of Physics, Tokyo Metropolitan University,\\ 1-1 Minami-Osawa, Hachioji, Tokyo 192-0397}
\altaffiltext{7}{Department of General Education, Kogakuin University,\\ 2665-1 Nakano-Cho, Hachioji, Tokyo 192-0015}  
\altaffiltext{8}{Kavli Institute for Astrophysics and Space Research,
Massachusetts Institute of Technology,\\ 77 Massachusetts Avenue,
Cambridge, MA 02139, USA}


\maketitle

\begin{abstract}
The charge transfer inefficiency (CTI) of the X-ray CCDs
on board the Suzaku satellite (X-ray Imaging Spectrometers;
XIS) has increased since the launch due to radiation damage,
and the energy resolution has been degraded.  To improve the
CTI, we have applied a spaced-row charge injection (SCI)
technique to the XIS in orbit; by injecting charges into CCD
rows periodically, the CTI is actively decreased. The CTI in the SCI 
mode depends on the distance between a
signal charge and a preceding injected row, and the
pulse height shows periodic positional variations. Using in-flight
data of onboard calibration sources and of the strong iron line
from the Perseus cluster of galaxies, we studied the
variation in detail. We developed a new method to correct the variation. 
By applying the new method, the energy resolution (FWHM) at 5.9~keV at March
2008 is $\sim$155~eV for the front-illuminated CCDs and $\sim$175~eV for
the back-illuminated CCD.

\end{abstract}

\section{Introduction}

X-ray charge coupled devices (CCDs) have good spatial and
energy resolution, and they have been the main detector for
imaging spectroscopy in X-ray astronomy since the ASCA
SIS~\citep{Bur93}.  X-ray CCDs in orbit, however, suffer
from radiation damage. The damage causes the increase of the
charge transfer inefficiency (CTI), which results in the 
degradation of the energy resolution for two
reasons: 1) the pulse height strongly depends on the position of an
X-ray event, since the X-ray event loses more electric
charges as the number of transfer increases, and 2) the loss
of charge is a stochastic process, and thus there is a fluctuation
in the amount of lost charge. In the case of the X-ray
Imaging Spectrometer (XIS; \cite{Koy07}) on board the Suzaku
satellite~\citep{Mit07}, the energy resolution at 5.9~keV
was $\sim$140~eV (FWHM) at August 2005, and had degraded to
$\sim$200~eV at August 2006.

The XIS is equipped with a charge injection (CI)
structure~\citep{Pri04,Bau04, Lam04,Pri08} which lies
adjacent to the top row of the imaging area. The CI
structure allows us to inject a commandable amount of charge
in a nearly arbitrary spatial pattern. We can measure the
CTI of each column precisely by the checker flag CI
technique, and it is possible to correct the lost charge.
The column-to-column CTI correction improves the energy
resolution greatly~\citep{Nak08, Oza09}.  However, we cannot
correct the fluctuation in principle, and the degraded
energy resolution cannot be fully restored even with the
column-to-column correction.

The CI can be used in another way to mitigate the effect of
the radiation damage; the spaced-row charge injection (SCI)
technique can reduce the CTI actively and improve the energy
resolution.  In the SCI technique, a charge is injected into
CCD rows periodically. The injected charge fills the
radiation-induced traps as a ``sacrificial charge", and thus
prevents some of the traps from capturing a signal charge
produced by X-rays.  Results based on ground experiments
using the SCI technique with radiation-damaged CCDs have
been reported~\citep{Tom97,Bau04}, but no in-orbit
experiment had been done. The Suzaku XIS operated with the SCI
technique in orbit for the first time at August 2006, and
the energy resolution was improved from $\sim$200~eV to
$\sim$140~eV at 5.9~keV~\citep{Bau07}. 
The SCI has been a normal observation mode since October 2006. 

In this paper, we report the study of the CTI in the SCI
mode, and present a new CTI correction method. Because the
CTI depends on the distance between a pixel and a
charge injected row, the calibration for the SCI mode
becomes complicated.  We developed a new method to correct
the complex CTI of the SCI mode to improve the
energy resolution further. 
We also found that
the CTI increased with time even with the SCI, 
and our method can correct the time
variation. All errors are at the 1$\sigma$
confidence level unless otherwise described.

\section{Spaced-row Charge Injection} 
 
In the SCI mode of the Suzaku XIS, a charge is injected into
every 54th row. The amounts of the injected charge into each
pixel are equivalent to the X-ray energy of $\sim$6~keV and
of $\sim$2~keV, for the front-illuminated (FI) and
back-illuminated CCD (BI), respectively.

The terminologies and notations for the CCD, CTI, and CI are
the same as \citet{Oza09}. To make clear for further
descriptions of the SCI, relevant terminologies and
notations are summarized in table~\ref{tab:notations}.

The CTI in the SCI mode is assumed to consist of two
components as CTI\,1 and CTI\,2 following \citet{Oza09}. The
relation among readout pulse height ($PH'$), original pulse
height generated by an X-ray of energy $E$ ($PH_{\rm o}$ at
$E$), CTI\,1 ($c_1$), CTI\,2 ($c_2$), and transfer number
$i$ are formalized in equation 2 of \citet{Oza09}, where the
detector coordinates are defined as $ActX$ and $ActY$.
The CCD consists of four segments (A, B, C, and D); each 
segment consists of 256 columns along $ActX$ and has 
a dedicated read-out node. 

In the SCI mode, a radiation-induced trap is filled with a
probability $p$ when a sacrificial charge passes through the
trap.  The filled trap does not capture a signal charge
produced by an X-ray.  However, the electron once filling
the trap will be re-emitted with the time scale $\tau$.
Thus the probability of the trap holding the electron is
proportional to $p \cdot \exp(-t/\tau)$, where $t$ is the
time elapsed since the sacrificial charge passes the trap.
Thus it is reasonable to assume that a pixel which
is $j$ rows away from its preceding charge injected row has
the $c_1$ value proportional to $1- p \cdot \exp(-\delta t
\cdot j/\tau)$, where $\delta t$ is time for one vertical
transfer; $\delta t \cdot j$ represents a time lag between
the pixel and the charge injected row.  We assume $\delta t
\cdot j /\tau$ is a small value, and hence $1- p \cdot
\exp(-\delta t \cdot j/\tau)$ is a linear function of $j$
approximately.  The charge is injected into every 54th row,
and hence $j= i \bmod 54$
\footnote{The modular arithmetic $i \bmod 54 = j$ means that when $i$
is divided by 54, it leaves $j$ as the remainder, hence
$i-j$ is divisible by 54.  For example, $111 \bmod
54=3$. Strictly speaking, charges are injected in rows of
$i=54 \cdot n -1$ $(n=0,\cdots, 18)$ and 1023 for full
window mode. Thus $j$ should be $ (i+1) \bmod 54$. We
nevertheless use this simplified description to avoid
unnecessary confusion. We treated $j$ properly for actual
CTI measurement.}.  We, consequently, can model the
$c_1$-$i$ relation of the SCI mode as,
\begin{eqnarray}
c_1(i) =  c_{1 \rm t} + \frac{c_{1\rm b} - c_{1\rm t}}{54} \cdot ( i  \bmod 54 ).
\label{eq:Q-ACTY}
\end{eqnarray}

\begin{figure*}
\FigureFile(69mm,60mm){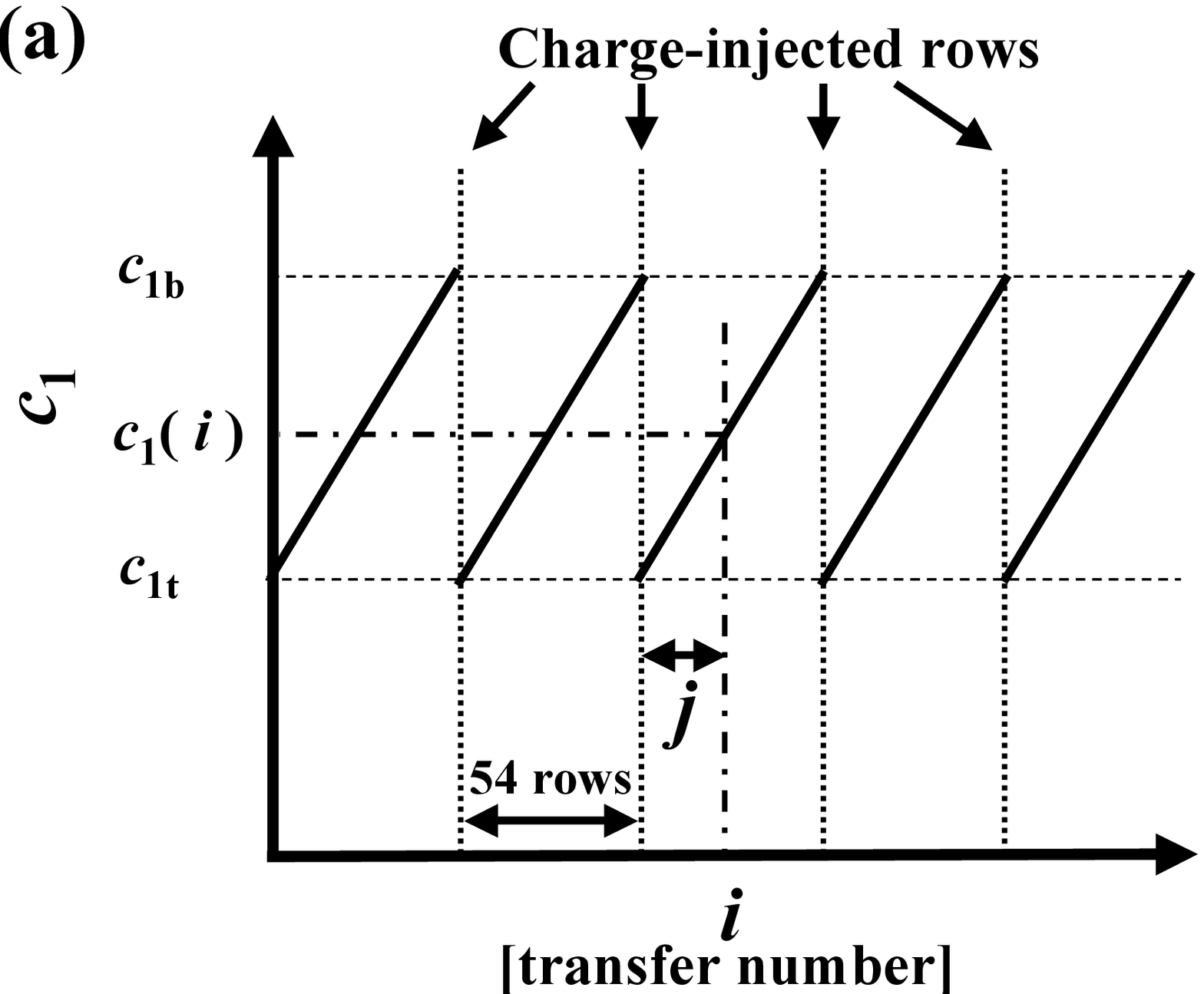}
\FigureFile(91mm,60mm){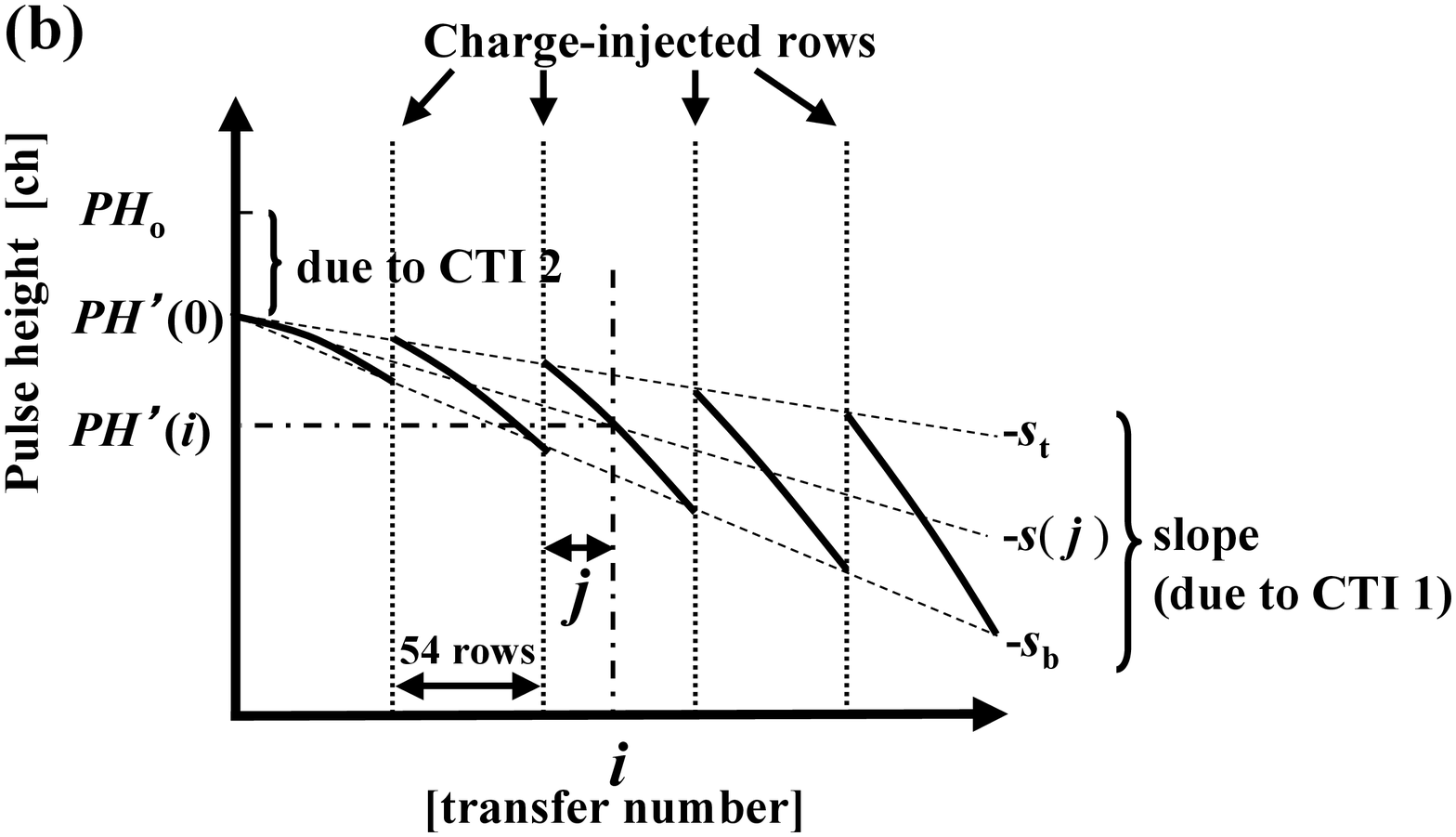}
\caption{ 
Our model of the ``sawtooth" variation. We assume the 
$c_1$-$i$ relation show in (a), and it generates the $PH'$-$i$ relation 
shown in (b).
\label{fig:saw-function} }
\end{figure*}

Equation \ref{eq:Q-ACTY} is a periodic sawtooth function as
demonstrated in figure~\ref{fig:saw-function}a. This
$c_1$-$i$ relation generates the $PH'$-$i$ relation in
figure~\ref{fig:saw-function}b, and the shape is nicely
reproduced by the ground experiments using the heavily
damaged CCD as demonstrated in figure 15 of \citet{Tom97}.

\begin{figure*}
\begin{center}
\FigureFile(80mm,60mm){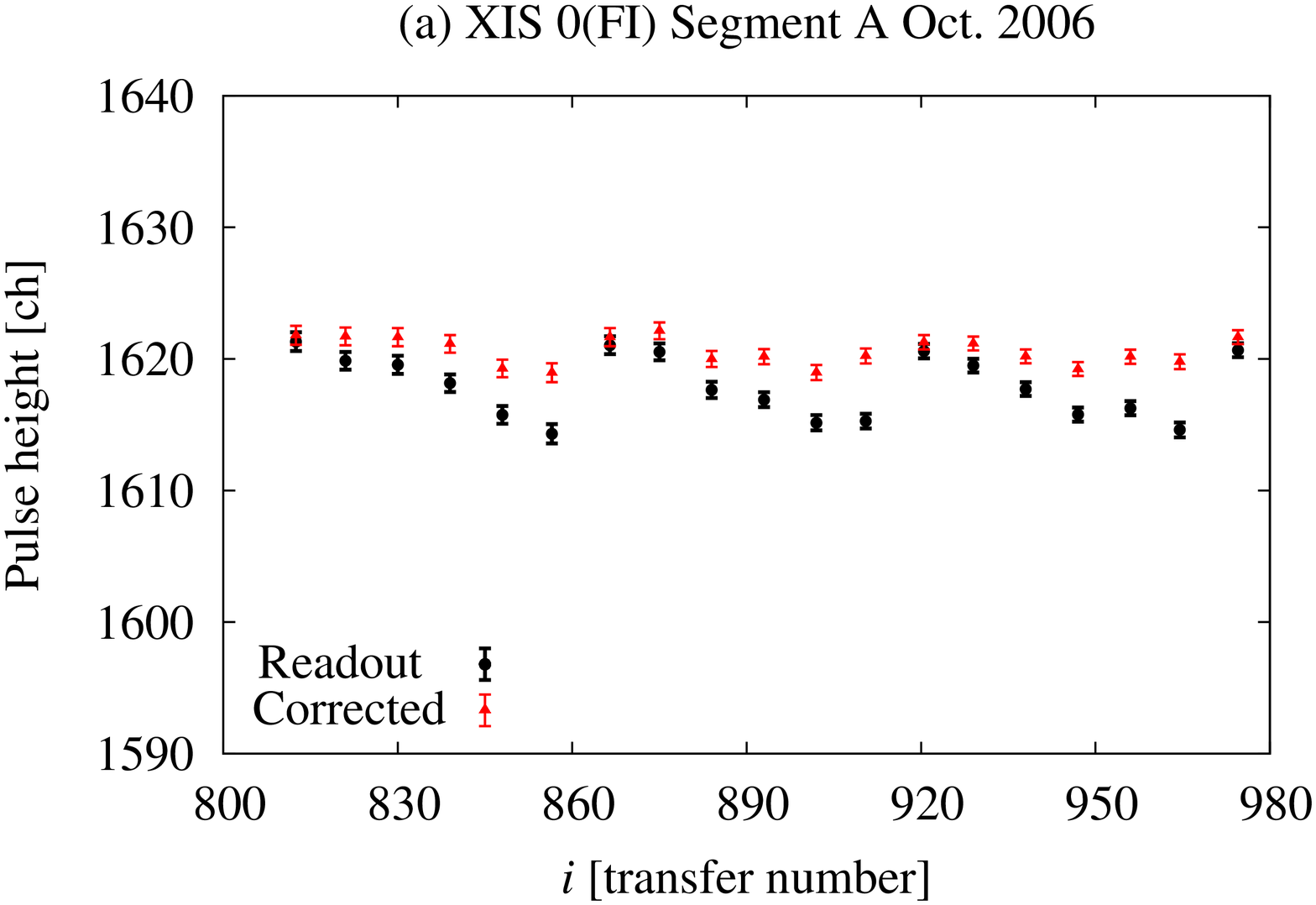}
\FigureFile(80mm,60mm){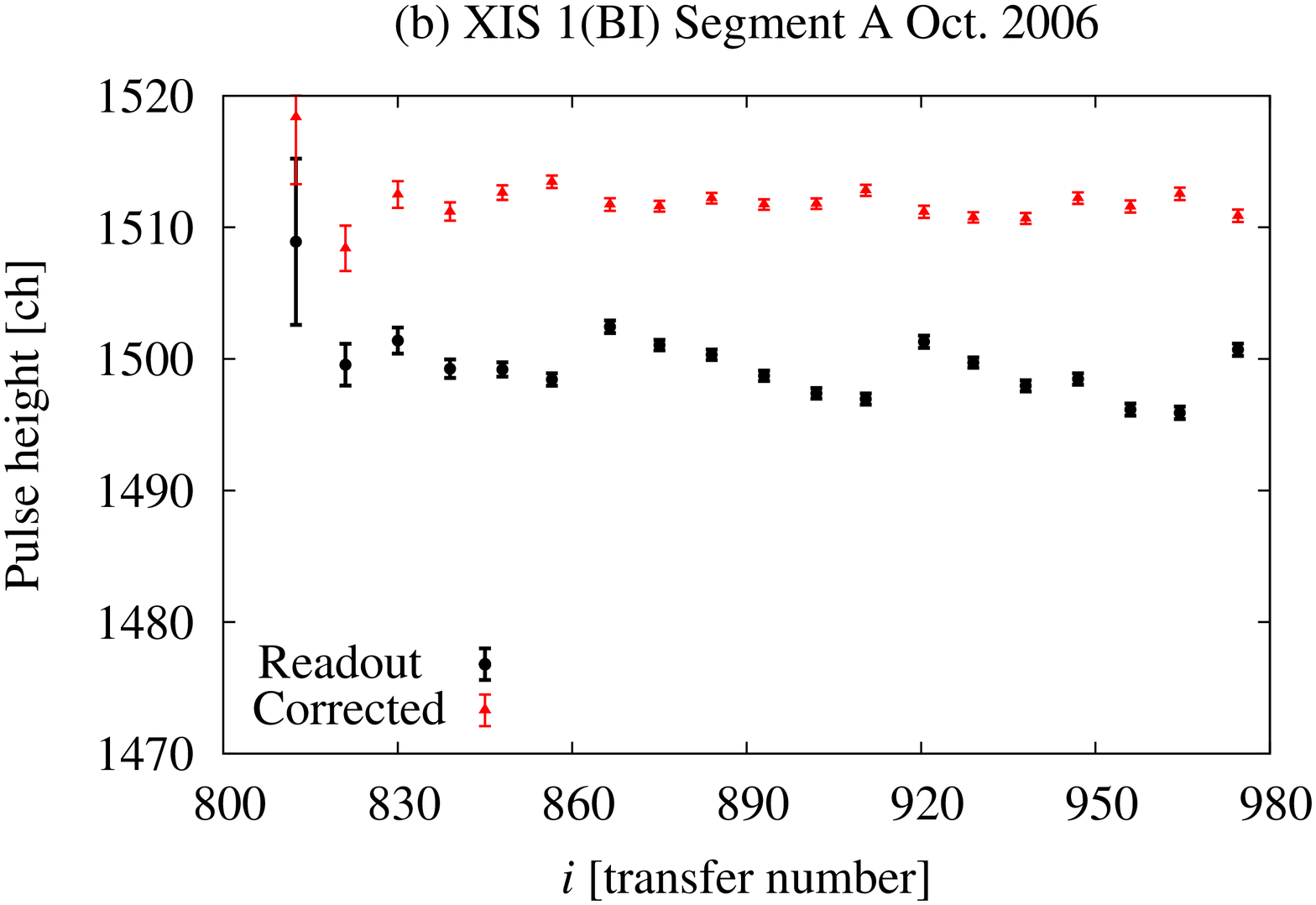}
\FigureFile(80mm,60mm){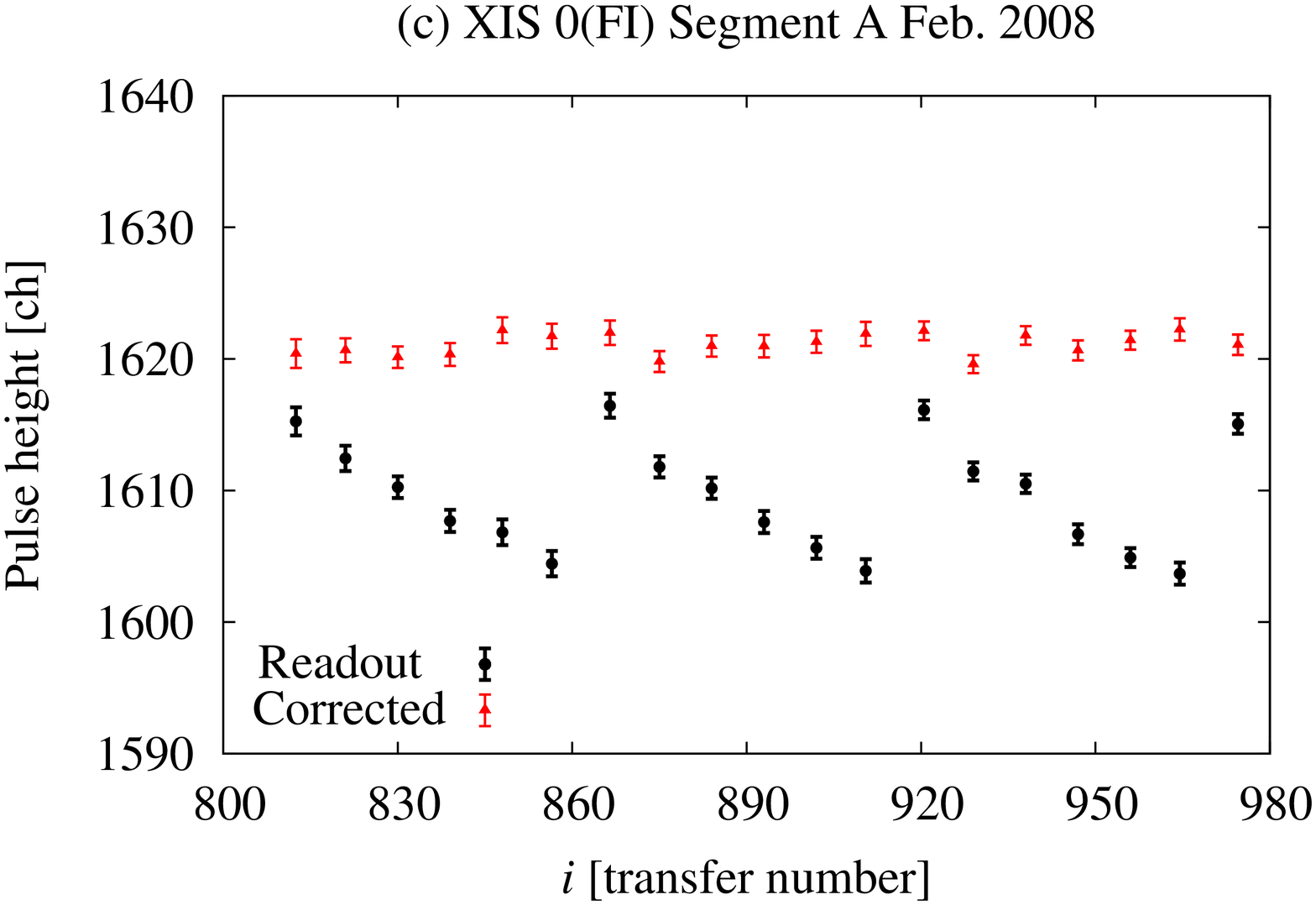}
\FigureFile(80mm,60mm){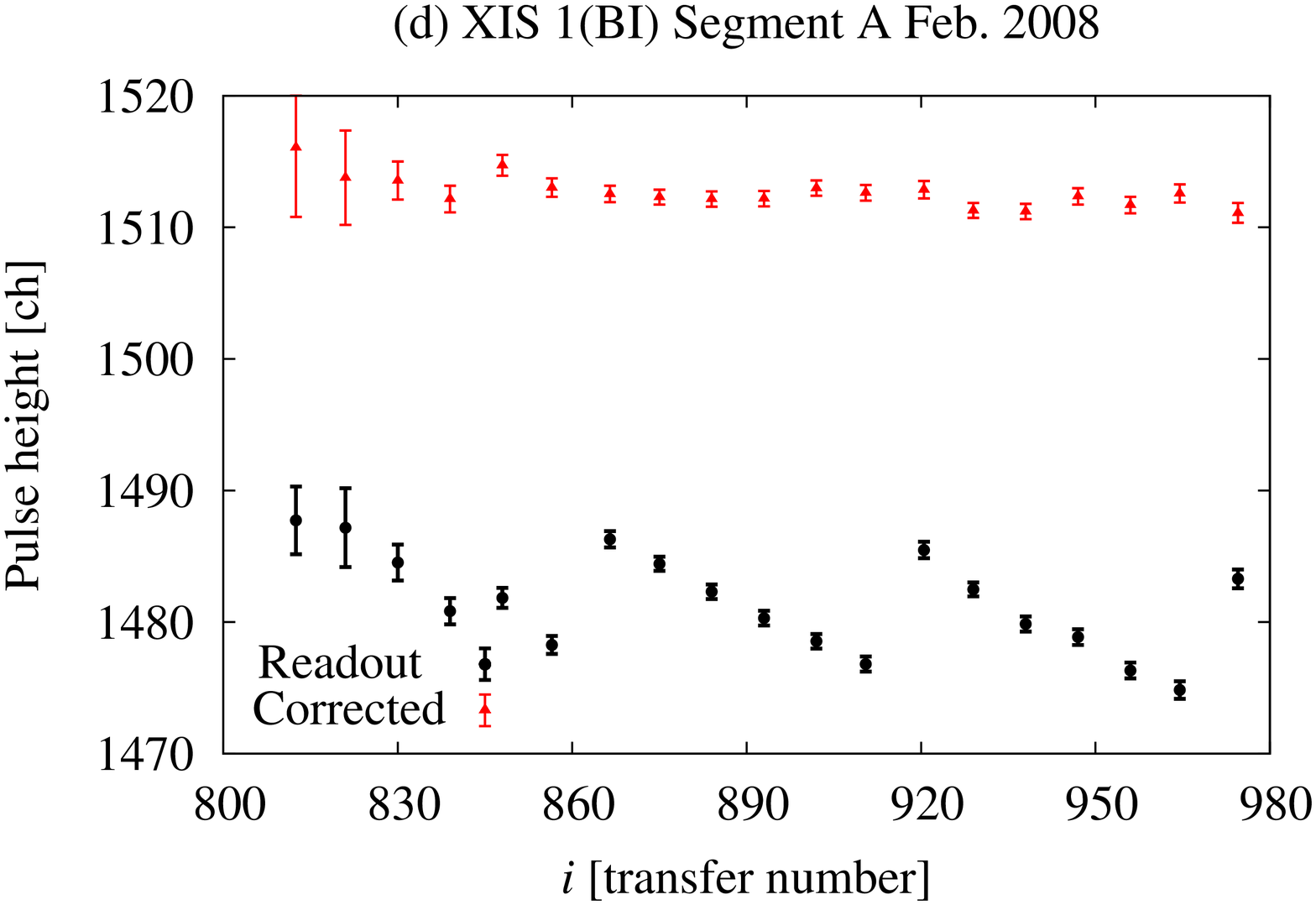}
\end{center}
\caption{
\label{fig:acty-pha} 
Pulse height of the Mn\emissiontype{I} K$\alpha$ line from the
onboard calibration source as a function of $i$:
October 2006 (a and b) and February 2008 (c and d).
We show the results of the segment A in XIS 0, 1 as typical examples. 
X-ray events of grade 02346 were analyzed. 
Black and red marks are data before and after our new CTI correction, respectively.  }
\end{figure*}

The sawtooth distribution is also found in the XIS in-orbit.
We measured the $PH'$-$i$ relations using the onboard
calibration source data in October 2006 with an effective
exposure time of 1~Ms, and show the results in
figures~\ref{fig:acty-pha}a and b.
Figure~\ref{fig:acty-pha} shows that the sawtooth of the BI
sensor is shallower than that of the FI sensors. The results
based on the data in February 2008 are also shown in
figures~\ref{fig:acty-pha}c and d. Compared with October
2006, the sawtooth became deeper. The pulse height just
after the charge injected rows changed more in the case of
the BI sensor than the FI sensors.  This difference might be
resulted in part from the smaller amount of charge injected
in the BI sensor.

Our sawtooth CTI model can represent the complicated
relation between $PH'$ and $i$, and make it possible to
convert $PH'(i)$ to $PH_{\rm o}$ with only three parameters,
$c_{1\rm t}$, $c_{1\rm b}$, and $c_2$.  Our goal is to
convert $PH^{\prime}(i)$ to $PH_{\rm o}$ by deciding the
three parameters.
 
\begin{table*}
\begin{center}
\caption{The list of notation.\label{tab:notations}}
\begin{tabular}{ll}	\hline 
Notation & Meaning \\	\hline
$i$ & Transfer number. $i$=$ActY+1$; $ActY$ is a coordinate value 
where an incident X-ray generates a charge.\\
$PH_{\rm o}$ & Original pulse height.\\
$PH'(0)$ & Readout pulse height of a pixel at $i=0$. It is equal to $(1- 1024 \cdot c_2)\cdot PH_{\rm o}$.\\
$PH'(i)$ & Readout pulse height of a pixel at $i$.\\
$j$ & Row number between a pixel and its preceding charge injected row. $j=i \bmod 54.$\\
$c_1(i)$ & CTI\,1 of  \citet{Oza09} for a pixel at $i$. \\
$c_{1 \rm t}$, $c_{1 \rm b}$ & The $c_1$ values at the peak and valley of the sawtooth (see figure 1a).\\
$c_2$ & CTI\,2 of \citet{Oza09}.\\
$-s(j)$ & Slope for a pixel with $j$. It is equal to $c_1(i) \cdot PH_{\rm o}$. \\
$-s_{\rm t}$ & Slope for the tops of the sawtooth. It is equal to $c_{1 \rm t} \cdot PH_{\rm o}$.\\
$-s_{\rm b}$ & Slope for the bottoms of the sawtooth.  It is equal to $c_{1 \rm b} \cdot PH_{\rm o}$.\\
$\beta$ & CTI depends on the pulse height as $c_{1,2} \propto PH_{\rm o}^{-\beta}$. See \citet{Oza09}.\\ \hline		
\end{tabular}
\end{center}
\end{table*}

\section{CTI Measurement in Orbit} 
\subsection{Calibration Data}

To study the CTI in the SCI mode, we analyzed the data of
the onboard calibration sources \atom{Fe}{}{55}, the Perseus
cluster of galaxies, and 1E~0102.2$-$7219, whose properties
are summarized in section~4 of \citet{Oza09}.

XIS 2 suddenly showed an anomaly on November 9, 2006, and it
has not been operated since then. Although there is no
direct evidence, the micro-meteoroid impact might have
caused the anomaly
\footnote{See http://www.astro.isas.jaxa.jp/suzaku/proposal/ao3/suzaku\_td/.}.
Thus only the data of XIS 0, 1 and 3 are studied.

All data were acquired with the normal clocking mode and the
full window option using the SCI. The editing mode was $3
\times 3$ or $5\times 5$. \citet{Koy07} provide details of
these modes. 

As mentioned in \citet{Koy07}, a small fraction of the 
charge in a pixel is left behind (trailed) to the next pixel 
during the transfer. This charge-trail phenomenon changes 
the spatial extent of an X-ray event. 
All data were corrected for the phenomenon based on the in-orbit data
\footnote{The details about the charge-trail correction 
is shown in\\
http://xmm2.esac.esa.int/external/xmm\_sw\_cal/icwg/\\presentations/Suzaku\_XIS.pdf
}, otherwise some X-ray events would be judged as grade 7,
and the detection efficiency would decrease.

We used the archival trend data of the calibration 
source obtained between August 2006 and
March 2008, and the total effective exposure time is about 23.7~Ms.
The observations of the celestial objects are summarized in
table~\ref{tab:obslog}.

\begin{table*}
\begin{center}
\caption{Log of Observations for Calibration.}\label{tab:obslog}
\begin{tabular}{cccc}	\hline 
Obs. ID	&\multicolumn{2}{c}{Observation time (UT)} & Exposure time\\
& Start & End & [ks]\\ \hline	
\multicolumn{4}{c}{The Perseus cluster} \\ \hline
101012010 & 2006/08/29 18:55:07 & 2006/09/02 01:54:19 &50.0 \\
101012020 & 2007/02/05 15:57:48 & 2007/02/06 14:30:14 &43.9 \\
102011010 & 2007/08/15 12:40:49 & 2007/08/16 11:27:22 &42.3 \\
102012010 & 2008/02/07 02:09:42 & 2008/02/08 10:30:19 &41.6 \\	\hline
\multicolumn{4}{c}{1E~0102.2$-$7219} \\ \hline
101005090 & 2006/12/13 18:53:16 & 2006/12/14 03:04:19 & 28.2 \\ 
101005110 & 2007/02/10 22:13:47 & 2007/02/11 19:30:14 & 36.2 \\ 
102001010 & 2007/04/10 10:35:08 & 2007/04/10 19:30:19 & 18.1 \\ 
102002010 & 2007/06/13 10:10:12 & 2007/06/14 03:31:19 & 27.9 \\ 
102003010 & 2007/08/12 05:21:09 & 2007/08/13 03:45:24 & 39.5 \\ 
102004010 & 2007/10/25 12:24:45 & 2007/10/26 09:00:14 & 26.2 \\ 
102005010 & 2007/12/01 19:25:40 & 2007/12/02 09:50:19 & 24.8 \\ 
102022010 & 2008/02/14 16:57:28 & 2008/02/16 03:10:24 & 26.5 \\ 
102006010 & 2008/03/15 05:43:27 & 2008/03/15 20:45:24 & 28.2 \\ 
\hline
\end{tabular}
\end{center}
\end{table*}

\subsection{Determination of the CTI Parameters}
The procedures for the CTI determination are as follows:
\begin{itemize}
\item {\it Step 1:} deciding $PH_{\rm o}$ at 5.895~keV and 6.56~keV.
\item {\it Step 2:} measuring $c_2$ for $PH_{\rm o}$ at 6.56~keV.
\item {\it Step 3:} measuring $c_{1 \rm t}$ and $c_{1 \rm b}$ for $PH_{\rm o}$ at 5.895~keV
or 6.56~keV.
\item {\it Step 4:} deciding the CTIs for any $PH'$ values.
\end{itemize}

In the case of the non-SCI mode, we can measure the CTI of
each column by the checker flag CI \citep{Nak08,Oza09}.
Since the checker flag CI is a complicated
operation, we have not used this technique in the SCI mode.
We, therefore, measured the averaged CTI of each segment.

In the normal analysis of the XIS data, both of single-pixel
and multi-pixel events (grade~0 and 2346 events; see
\cite{Koy07}) are used. To determine the CTI, however, we
analyzed only the grade~0 events; if we use the multi-pixel
events, it is difficult to measure the CTI correctly,
because the CTI depends on the amount of transferred charge,
and the amounts of charge in each pixel comprising the
multi-pixel event is different with each other.  We will
mention how to correct the data of grade~02346 events in
step 4.

\subsection*{Step 1: $PH_{\rm o}$ at 5.895~keV and 6.56~keV}

We determined the $PH_{\rm o}$s of the
Fe\emissiontype{XXV} K$\alpha$ line from the Perseus cluster
(6.56~keV) by using the data of the first SCI observation on 29 August, 2006. 
Because it is impossible to measure the $c_2$ value in August 
2006, we assumed that value is equal to zero. Even if this 
assumption would not be reasonable, we can cover it 
by adjusting the  $PH_{\rm o}$-$E$ relation. Then the
$PH'$ at $i=0$ ($PH'(0)$) is equal to $PH_{\rm o}$.

\begin{figure}
\begin{center}

\FigureFile(80mm,60mm){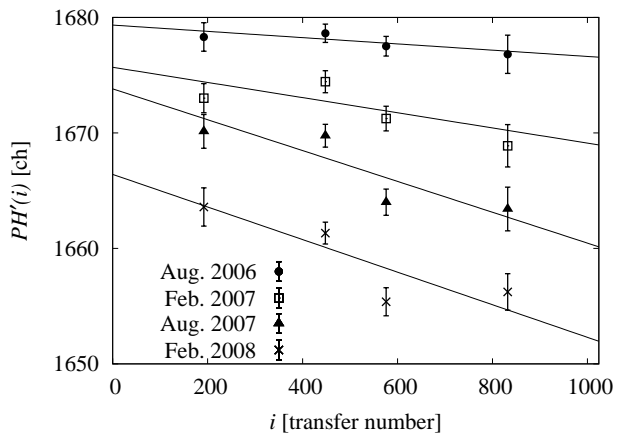}
\end{center}
\caption{
\label{fig:CTI2} 
$PH'$ as a function of $i$ obtained using the
Fe\emissiontype{XXV} K$\alpha$ line from the Perseus cluster.
The result of the XIS~1 segment C is shown as a typical example.
}
\end{figure}

Since the statistics are limited, we divided each segment to
four regions along the $ActY$ axis, and obtained the center
pulse height of the iron line from each region.  An example
of $PH'$ as a function of $i$ is shown in
figure~\ref{fig:CTI2}. By fitting the data with a linear
function of $i$, the $PH_{\rm o}$ value was obtained as 
$PH'(0)$

For the segments A and D, we obtained the $PH_{\rm o}$ value
of the Mn \emissiontype{I} K$\alpha$ line (5.895~keV) using
the data of the calibration sources in August, 2006. We
obtained the $PH'$-$i$ relation, but in this case, the $i$
values are limited at around 900 because the 
calibration sources irradiate only the two far-end corners 
from the read-out node of the imaging area \citep{Koy07}.
We extrapolated the relation to $i=0$ by using a linear function, the slope of
which was fixed to a value $s_{\rm Mn}$ calculated as follows.  Because
the slope represents an averaged $-c_1\cdot PH_{\rm o}$, the
slope should be proportional to $PH_{\rm o}^{1-\beta}$. Then
we calculated $s_{\rm Mn}$ from the slope for the 6.56~keV data
($s_{\rm Fe}$) as $s_{\rm Mn} = s_{\rm Fe} \cdot
(5.895/6.56)^{1-\beta} = s_{\rm Fe}\cdot(0.90)^{1-\beta}$.

\subsection*{Step 2: $c_2$ for $PH_{\rm o}$ at 6.56~keV}

We obtained the $PH'(0)$ values for all data of the Perseus
cluster (table~2) with the same method of
step 1. The $PH'$-$i$ relations are shown in
figure~\ref{fig:CTI2}.  As is found in
figure~\ref{fig:CTI2}, $PH'(0)$ is decreased with time;
$PH'(0)$ becomes different from $PH_{\rm o}$. We assumed the
difference $PH_{\rm o} - PH'(0)$ is attributable to $c_2$.
Assuming that $c_2$ is a linear function of time, the
increasing rate of $c_2$ for $PH_{\rm o}$ (at 6.56~keV) is
typically $\sim 2
\times 10^{-6}$~yr$^{-1}$ for the FI sensors and $\sim 6
\times 10^{-6}$~yr$^{-1}$ for the BI sensor.

\subsection*{Step 3: $c_{1 \rm t}$ and $c_{1 \rm b}$ for $PH_{\rm o}$ at 5.895~keV and 6.56~keV}

\begin{figure}
\begin{center}
\FigureFile(80mm,60mm){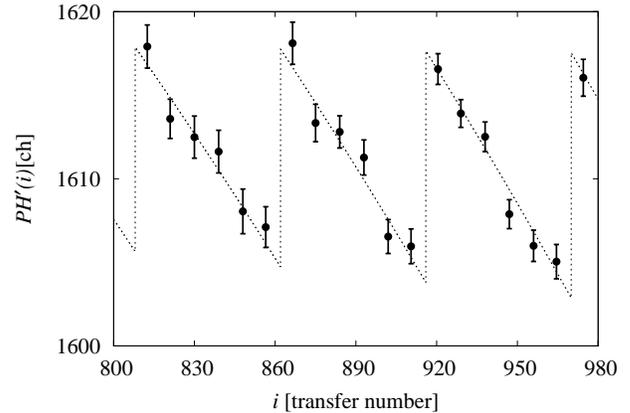}
\end{center}
\caption{
\label{fig:CTI1_cal} 
Best-fitting result of the sawtooth function for
the Mn\emissiontype{I} K$\alpha$ line of the calibration
source. The result of the XIS~0 segment A in February 2008
is shown as a typical example. Only grade 0 events were
used.  }
\end{figure}

\begin{figure}
\begin{center}
\FigureFile(80mm,60mm){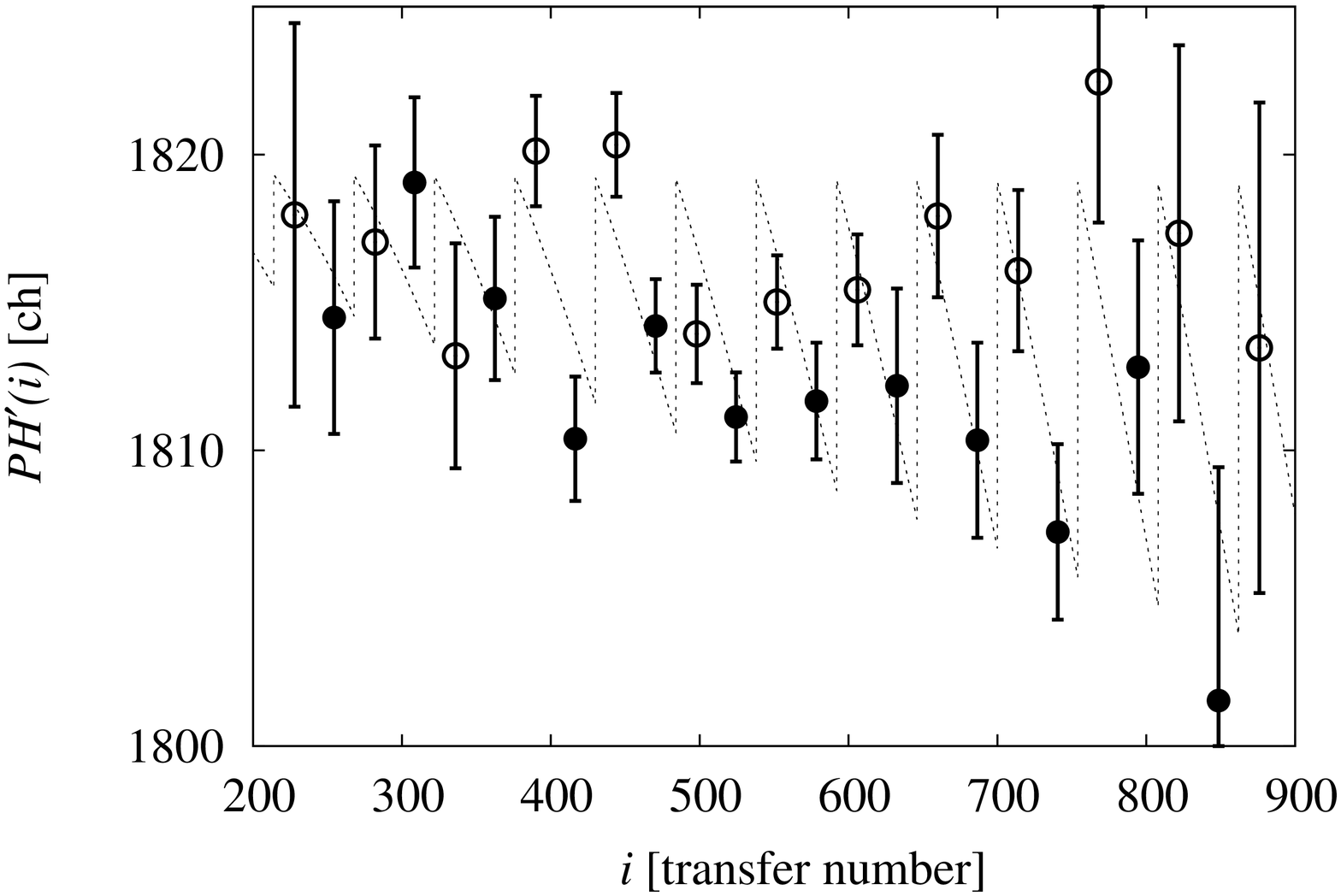}
\end{center}
\caption{
\label{fig:CTI1_A426} 
Best-fitting result of the sawtooth function for the
Fe\emissiontype{XXV} K$\alpha$ line from the Perseus
cluster. The result of the XIS~0 segment C in February 2008 is
shown as a typical example.  The white and black circles
show the pulse height of regions with smaller and larger $j$
values, respectively.  Only grade 0 events were used.  }
\end{figure}

For the segments A and D, we determined $s_{\rm t}$ and
$s_{\rm b}$ for $PH_{\rm o}$ at 5.895~keV by using the data of
the calibration sources. For the segments B and C, we
measured those for $PH_{\rm o}$ at 6.56~keV by using the
Perseus data.  We fitted the $PH'$-$i$ relation with
the sawtooth function shown in figure~\ref{fig:saw-function}b, 
and determined $s_{\rm t}$ and
$s_{\rm b}$.  In the fitting, we fixed the $PH_{\rm o}$ and
$c_2$ to the values obtained in step 1 and step 2.
Typical examples of the fitting are shown in
figures~\ref{fig:CTI1_cal} and \ref{fig:CTI1_A426}.  Using 
$s_{\rm t}$ and $s_{\rm b}$, we calculated $c_{1 \rm t}$ and $c_{1 \rm b}$.

\begin{figure}
\begin{center}
\FigureFile(80mm,60mm){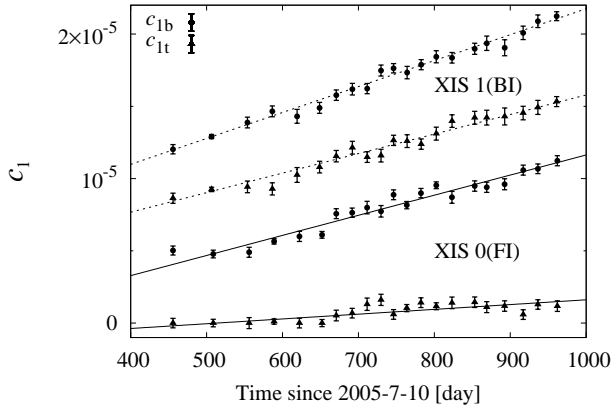}
\end{center}
\caption{
\label{fig:TimeHisCTI1} 
Time history of $c_{1 \rm t}$ and $c_{1 \rm b}$ for $PH_{\rm o}$ at 5.895~keV.
The segment A of XIS~0 and 1 are shown as typical examples. 
}
\end{figure}

In the case of the FI sensors, the $c_{1 \rm t}$ and $c_{1
\rm b}$ values for $PH_{\rm o}$ at 5.895~keV at August 2006 are
typically $< 10^{-7}$ and $\sim 3 \times 10^{-6}$,
respectively. The increasing rates are $\sim 1\times
10^{-6}$~yr$^{-1}$ and $\sim 5 \times 10^{-6}$~yr$^{-1}$,
respectively. In the case of the BI sensor, the parameters
$c_{1 \rm t}$ and $c_{1 \rm b}$ at August 2006 are $\sim 6
\times 10^{-6}$ and $\sim 1 \times 10^{-5}$ , and the
increasing rates are $\sim 5 \times 10^{-6}$~yr$^{-1}$ and
$\sim 8 \times 10^{-6}$~yr$^{-1}$, respectively.  Typical
examples of the time evolution of the $c_1$ values for
$PH_{\rm o}$ at 5.895~keV are shown in
figure~\ref{fig:TimeHisCTI1}.

The CTI of the BI sensor is larger, and increases more
rapidly than that of the FI sensors.  It might be the reason
that the amount of the injected charge to the BI sensor is
less than that of FIs.

\subsection*{Step 4: CTI for any $PH'$ values}

For an event of $PH'$, we calculated the CTI with an
equation of $c_{1,2}
\times \{PH' / (PH_{\rm o} {\rm~at~5.895~keV~or~6.56~keV})\}^{-\beta}$.
Here, we assumed the same $PH_{\rm o}$-$c_{1,2}$ relation as
that of the non-SCI mode and used the same $\beta$
(typically $\sim 0.25$; \cite{Oza09}).

\begin{figure}
\begin{center}
\FigureFile(80mm,60mm){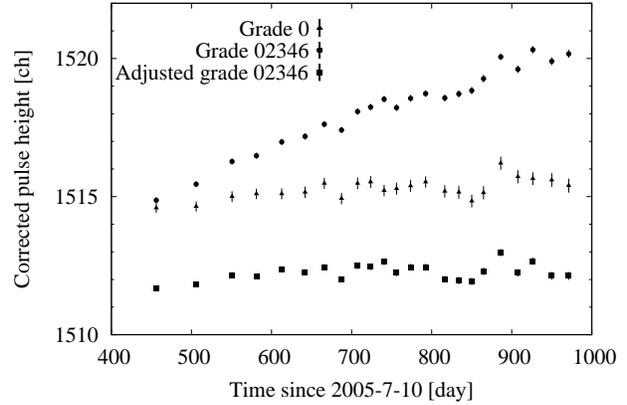}
\end{center}
\caption{
\label{fig:G0vs02346} 
Time history of the line centroid of Mn\emissiontype{I}
K$\alpha$ line. We show the result of the XIS~1 segment A 
as a typical example. 
The triangle and circle marks show the data of
grade~0 and 02346 after the CTI correction with the
parameters based on only the grade~0 events.  
The rectangle marks show the data of grade 02346 events 
after the CTI correction
with the fine-tuned parameters as described in the text.
 }
\end{figure}

Since we obtained the CTI for all pulse-height values, we
can now correct the CTI of multi-pixel events (grade~2346
events).  By using the Mn\emissiontype{I} K$\alpha$ line of
the calibration sources or the Fe\emissiontype{XXV}
K$\alpha$ line from the Perseus cluster, we examined the CTI
correction for the grade~02346
events. Figure~\ref{fig:G0vs02346} shows the line centroid 
of the grade~0 and grade~02346 events as a function
of time.  While the line centroid of the grade~0 events is
temporally constant, that of the grade~02346 events
increases with time.  This means that the CTI parameters for
the grade 2346 events are smaller than those for the grade 0
events. The CTI of the grade 2346 events can be smaller,
because the split charge acts as the sacrificial charge.

We then fine-tuned the parameters $c_{1 \rm t}$, $c_{1 \rm
b}$, and $c_2$ by multiplying a common time-independent
factor. We determined the factor of each segment so that the
line center of the grade~02346 events becomes temporally
constant. The factors are typically 0.9.
Figure~\ref{fig:G0vs02346} also shows that the result of the
grade~02346 data corrected with the fine-tuned CTI
parameters. The adjusted pulse height becomes constant.
However, as a result, the value becomes different from
$PH_{\rm o}$ measured in step 1. We regarded
this adjusted pulse height as $PH_{\rm o}$ at 5.895~keV, and
determined the $PH_{\rm o}$-$E$ relation.

\begin{figure}
\begin{center}
\FigureFile(80mm,60mm){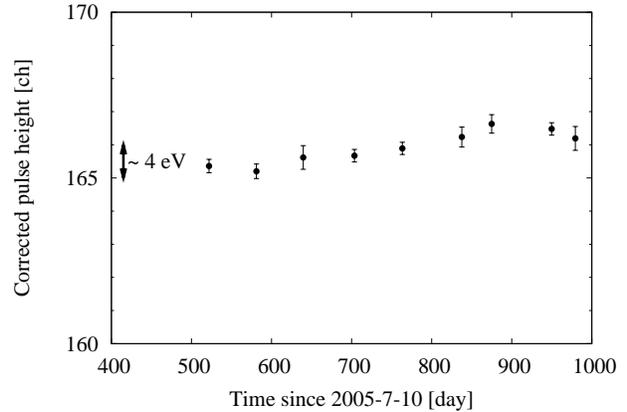}
\end{center}
\caption{
\label{fig:E0102} 
Time history of the line centroid of O\emissiontype{VIII} K$\alpha$ (0.653~keV). 
We show the result of the XIS~1 segment C as a typical example.
The results are based on the grade 02346 events corrected with 
the fine-tuned CTI parameters as described in the text.
}
\end{figure}

To check the CTI correction in the low-energy band, we
applied the fine-tuned CTI parameters to the grade~02346
data of 1E~0102.2$-$7219.  We fitted the spectrum with the
empirical calibration model\footnote{See
http://cxc.harvard.edu/acis/E0102/}.  Figure~\ref{fig:E0102}
shows the pulse height of the O\emissiontype{VIII} K$\alpha$
line after the CTI correction.  We can see that the line
centroid is constant, which supports the validity of our
method.

\section{Energy Scale Uniformity and Resolution in the SCI Mode}      

The pulse height of the Mn\emissiontype{I} K$\alpha$ line after our new CTI
correction is shown in figure~\ref{fig:acty-pha}.  The sawtooth structure disappeared, 
and hence our new method greatly
reduces the variation of the pulse height.  Comparing between
October 2006 and February 2008 shows that the CTI variation
with time is also corrected properly.

The $PH_{\rm o}$-$E$ relation for each segment is assumed to be the same
function form with a similar fine-tuning process of the absolute energy as \citet{Oza09}: 
a model of two slopes crossing at the energy of the Si-K edge (1.839~keV)  
with the same ratio of the two slopes as that  obtained in the 
ground experiments~\citep{Koy07}. 

\begin{figure}
\begin{center}
\FigureFile(80mm,60mm){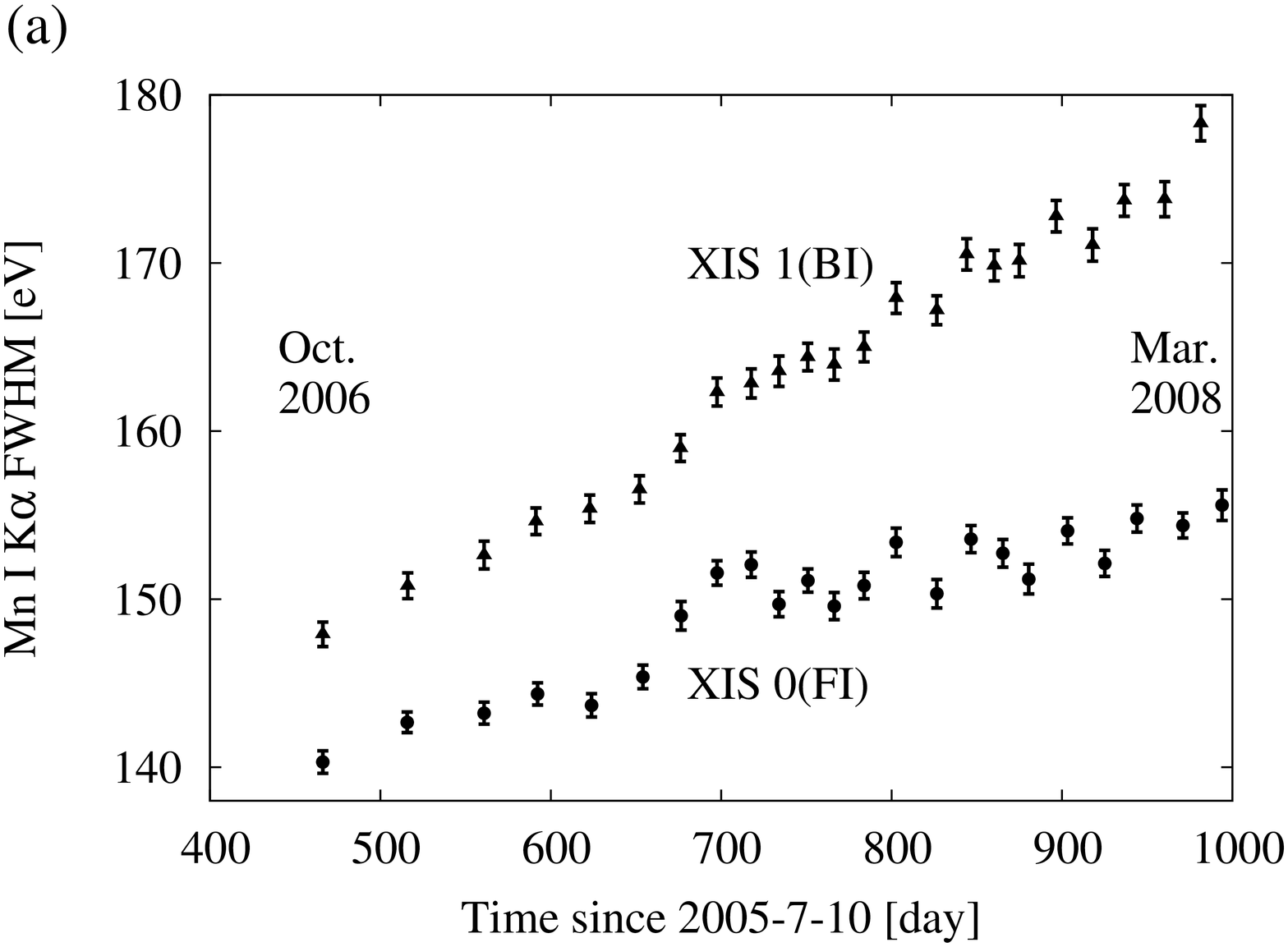}
\FigureFile(80mm,60mm){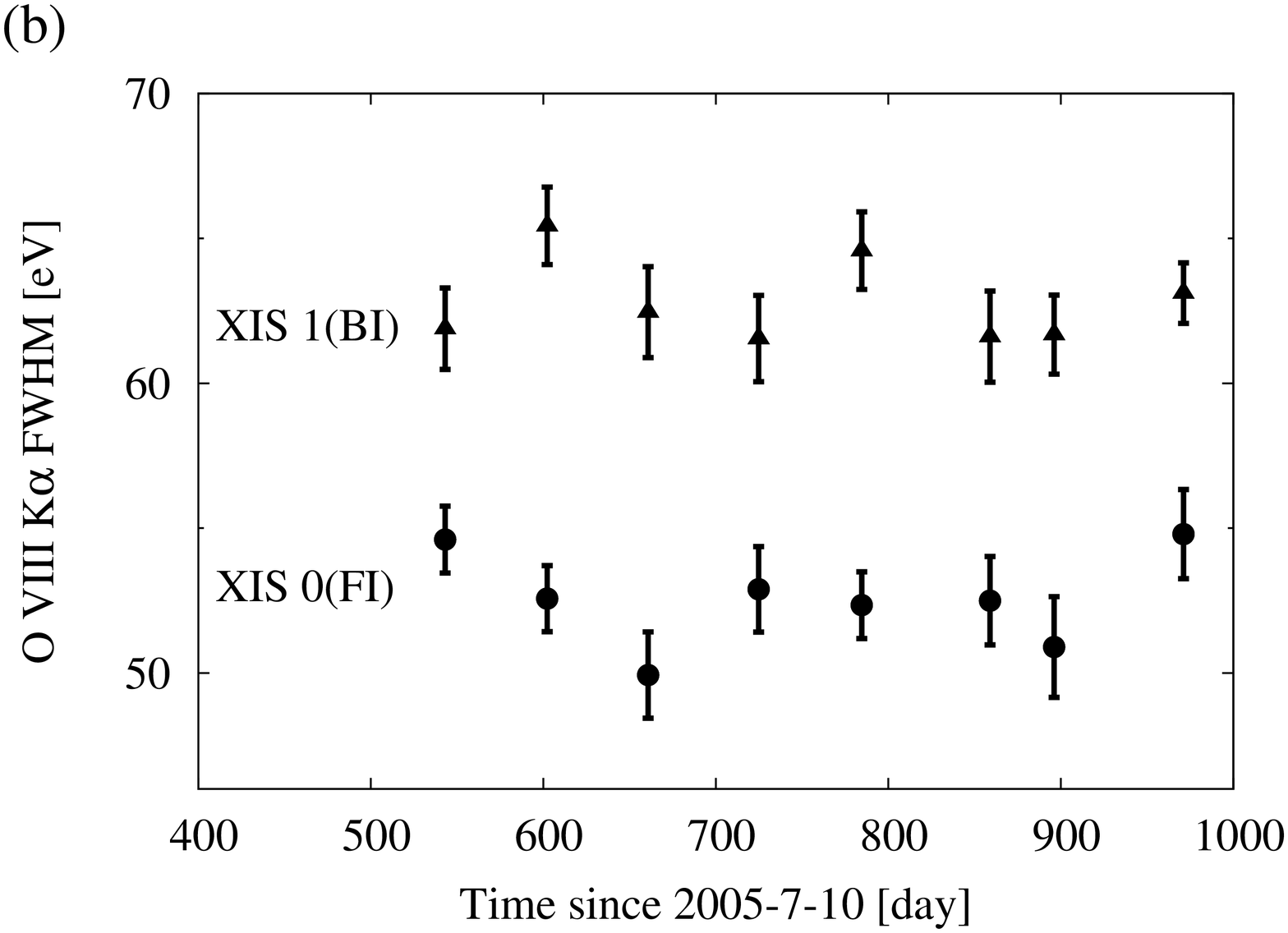}
\end{center}
\caption{
\label{fig:FWHM} 
Time history of the energy resolution (FWHM) at Mn\emissiontype{I} K$\alpha$ (5.895~keV) (a) 
and at O\emissiontype{VIII} K$\alpha$ (0.653~keV) (b). The
averaged value of the segments A and D are shown for each sensor. 
The results are based on the grade 02346 events corrected with 
the fine-tuned CTI parameters as described in the text.}
\end{figure}

The time histories of the energy resolution of the corrected
data are shown in figure~\ref{fig:FWHM}. At the high energy
(5.895~keV), the energy resolution of XIS 0 at March 2008 is
improved from $\sim$160~eV to $\sim$155~eV by the sawtooth
correction. On the other hand, the energy resolution of
XIS~1 at March 2008 is not improved in spite of the sawtooth correction,
and stays at $\sim$175~eV.
At the low energy (0.653~keV), no clear effect of the sawtooth correction 
is seen in both FI and BI.   
The energy resolutions at March 2008 
are $\sim$53~eV (XIS~0) and $\sim$62~eV (XIS~1)
independently of the correction. 

As we mentioned in section 2, the sawtooth of the BI 
sensor is shallower than that of the FI sensors. We think it 
causes the difference of the effect of the sawtooth correction 
between BI and FI. 
We speculate, at the low energy, the effect of the sawtooth 
correction was not seen because the original degradation 
of the energy resolution was small. In fact, the energy 
resolutions at O\emissiontype{VIII} K$\alpha$ are almost 
same between the SCI and non-SCI mode. 

The results of our new method to correct the sawtooth
structure has been implemented to the software package
released by HEASARC\footnote{http://heasarc.gsfc.nasa.gov/}
since HEAsoft version 6.3. Now, all of the XIS data of the
SCI mode after the processing version 2.0 are corrected by
the sawtooth method.

\bigskip

We thank all the Suzaku team members for their support of the
observation and useful information on the XIS
calibration. Thanks Dr. Junko Hiraga of RIKEN for her beneficial comments. 
H.U., M.O., H.U., H.N., A.B., and D.T. are supported by JSPS Research
Fellowships for Young Scientists.  H.M. is supported by the
MEXT, Grant-in-Aid for Young Scientists~(B), 18740105, 2008,
and is also supported by the Sumitomo Foundation, Grant for
Basic Science Research Projects, 071251, 2007. 
H.T. and K.H. were supported by the MEXT, Grant-in-Aid 16002004.
This work was supported by the Grant-in-Aid for the Global COE Program ``The 
Next Generation of Physics, Spun from Universality and Emergence" 
from the Ministry of Education, Culture, Sports, Science and 
Technology (MEXT) of Japan.


\end{document}